\documentclass[usegraphicx,usenatbib]{mn2e}
\usepackage{amsmath}
\usepackage{amssymb}
\usepackage{color}
\usepackage{url}

\usepackage{threeparttable}

\newcommand\lsim{\mathrel{\rlap{\lower4pt\hbox{\hskip1pt$\sim$}}
        \raise1pt\hbox{$<$}}}
\newcommand\gsim{\mathrel{\rlap{\lower4pt\hbox{\hskip1pt$\sim$}}
        \raise1pt\hbox{$>$}}}
\newcommand{\lya}{\ifmmode\mathrm{Ly}\alpha\else{}Ly$\alpha$\fi}
\newcommand{\lyb}{\ifmmode\mathrm{Ly}\beta\else{}Ly$\beta$\fi}
\newcommand{\igm}{\ifmmode\mathrm{IGM}\else{}IGM\fi}
\newcommand{\lae}{\ifmmode\mathrm{LAE}\else{}LAE\fi}
\newcommand{\h}{\ifmmode\mathrm{H}\else{}H\fi}
\newcommand{\hi}{\ifmmode\mathrm{H\,{\scriptscriptstyle I}}\else{}H\,{\scriptsize I}\fi}
\newcommand{\hii}{\ifmmode\mathrm{H\,{\scriptscriptstyle II}}\else{}H\,{\scriptsize II}\fi}
\newcommand{\cmb}{\ifmmode\mathrm{CMB}\else{}CMB\fi}
\newcommand{\qso}{\ifmmode\mathrm{QSO}\else{}QSO\fi}
\newcommand{\eor}{\ifmmode\mathrm{EoR}\else{}EoR\fi}
\newcommand{\heii}{\ifmmode\mathrm{He\,{\scriptscriptstyle II}}\else{}He\,{\scriptsize II}\fi}
\newcommand{\heiii}{\ifmmode\mathrm{He\,{\scriptscriptstyle III}}\else{}He\,{\scriptsize III}\fi}
\newcommand{\ciii}{\ifmmode\mathrm{C\,{\scriptscriptstyle III]}}\else{}C\,{\scriptsize III]}\fi}
\newcommand{\oiii}{\ifmmode\mathrm{O\,{\scriptscriptstyle III}}\else{}O\,{\scriptsize III}\fi}
\newcommand{\aliii}{\ifmmode\mathrm{Al\,{\scriptscriptstyle III}}\else{}Al\,{\scriptsize III}\fi}
\newcommand{\mgii}{\ifmmode\mathrm{Mg\,{\scriptscriptstyle II}}\else{}Mg\,{\scriptsize II}\fi}
\newcommand{\fe}{\ifmmode\mathrm{Fe}\else{}Fe\fi}
\newcommand{\nv}{\ifmmode\mathrm{N\,{\scriptscriptstyle V}}\else{}N\,{\scriptsize V}\fi}
\newcommand{\niv}{\ifmmode\mathrm{N\,{\scriptscriptstyle IV]}}\else{}N\,{\scriptsize IV]}\fi}
\newcommand{\cii}{\ifmmode\mathrm{C\,{\scriptscriptstyle II}}\else{}C\,{\scriptsize II}\fi}
\newcommand{\civ}{\ifmmode\mathrm{C\,{\scriptscriptstyle IV}}\else{}C\,{\scriptsize IV}\fi}
\newcommand{\siv}{\ifmmode\mathrm{Si\,{\scriptscriptstyle IV}}\else{}Si\,{\scriptsize IV}\fi}
\newcommand{\siii}{\ifmmode\mathrm{Si\,{\scriptscriptstyle II}}\else{}Si\,{\scriptsize II}\fi}
\newcommand{\siiii}{\ifmmode\mathrm{Si\,{\scriptscriptstyle III]}}\else{}Si\,{\scriptsize III]}\fi}
\newcommand{\ovi}{\ifmmode\mathrm{O\,{\scriptscriptstyle VI}}\else{}O\,{\scriptsize VI}\fi}
\newcommand{\sioiv}{\ifmmode\mathrm{Si\,{\scriptscriptstyle IV}\,\plus O\,{\scriptscriptstyle IV]}}\else{}Si\,{\scriptsize IV}\,+O\,{\scriptsize IV]}\fi}
\newcommand{\Msun}{M_\odot}

\newcommand{\nf}{x_{\rm HI}}
\newcommand{\avenf}{$\bar{x}_{\rm HI}$}

\newcommand{\smallHII}{\textsc{\small Small \hii{}}}
\newcommand{\intermediateHII}{\textsc{\small Intermediate \hii{}}}
\newcommand{\largeHII}{\textsc{\small Large \hii{}}}

\pdfoutput=1
\title[Reionisation from ULASJ1342+0928]{Constraints on reionisation from the $\bmath{z=7.5}$ QSO ULASJ1342+0928}

\author[B. Greig et al.] {Bradley~Greig$^{1,2}$\thanks{E-mail:~greigb@unimelb.edu.au}, Andrei~Mesinger$^{3}$, \& Eduardo Ba{\~n}ados$^{4}$\\
$^1$ARC Centre of Excellence for All-Sky Astrophysics in 3 Dimensions (ASTRO 3D), University of Melbourne, VIC 3010, Australia \\
$^2$School of Physics, University of Melbourne, Parkville, VIC 3010, Australia \\
$^3$Scuola Normale Superiore, Piazza dei Cavalieri 7, I-56126 Pisa, Italy \\
$^4$The Observatories of the Carnegie Institution for Science, 813 Santa Barbara Street, Pasadena, California 91101, USA
}

\begin{document}
\maketitle \begin{abstract}
\noindent
The recent detection of ULASJ1342+0928, a bright QSO at $z=7.54$, provides a powerful probe of the ionisation state of the intervening intergalactic medium, potentially allowing us to set strong constraints on the epoch of reionisation (EoR).
Here we quantify the presence of Ly$\alpha$ damping wing absorption from the EoR in the spectrum of ULASJ1342+0928.
Our Bayesian framework simultaneously accounts for uncertainties on: (i) the intrinsic QSO emission (obtained from reconstructing the \lya\ profile from a covariance matrix of emission lines) and (ii) the distribution of \hii{} regions during reionisation (obtained from three different 1.6$^3$ Gpc$^3$ simulations spanning the range of plausible EoR morphologies). Our analysis is complementary to that in the discovery paper (Ba{\~n}ados et al.) and the accompanying method paper (Davies et al.) as it focuses solely on the damping wing imprint redward of \lya\ ($1218 < \lambda < 1230$\AA), and uses a different methodology for (i) and (ii).
We recover weak evidence for damping wing absorption.  Our intermediate EoR model yields the following constraints on the volume-weighted neutral hydrogen fraction at $z=7.5$: $\bar{x}_{\hi{}} = 0.21\substack{+0.17 \\ -0.19}$ (68 per cent).  The constraints depend weakly on the EoR morphology.
Our limits are lower than those presented by Ba{\~n}ados et al. and Davies et al., though they are consistent at $\sim$ 1 -- 1.5 $\sigma$.
We attribute this difference to: (i) a lower amplitude intrinsic \lya\ profile obtained from our reconstruction pipeline, driven by correlations with other high-ionisation lines in the spectrum which are relatively weak; and 
(ii) only considering transmission redward of Ly$\alpha$ when computing the likelihood, which reduces the available constraining power but makes the results less model-dependent. 
Our results are consistent with previous estimates of the EoR history, and support the picture of a moderately extended EoR.
\end{abstract} 

\section{Introduction}

The epoch of reionisation (\eor) denotes the final major baryonic phase change in the Universe; when the pervasive, dense neutral hydrogen fog is lifted by the cumulative ionising radiation from the first stars, galaxies and QSOs. The timing and duration of the EoR can loosely be constrained from indirect measurements such as the integral constraints on the reionisation history (\hii{} fraction) from the Thomson scattering of photons \citep[e.g.][]{George:2015p5869,Collaboration:2015p4320}. More direct, though often controversial, constraints on the latter stages of the EoR can be made from the absorption of \lya\ photons by lingering cosmic \hi{} patches. However, the intergalactic medium (IGM) at $z\gtrsim 6$ becomes dense enough that even if a small fraction of hydrogen is neutral ($\nf \gsim 10^{-4}-10^{-5}$), the vast majority of photons which redshift into \lya\ resonance are absorbed.  Thus the \lya\ forest saturates at high-$z$ \citep{Fan:2006p4005}.

A more versatile probe of the IGM neutral fraction is the \lya\ damping wing (e.g. \citealt{Rybicki1979,MiraldaEscude:1998p1041}).
These Lorenzian wings of the \lya\ profile are extended and relatively smooth functions of frequency.  The absorption cross-section in these wings is reduced by $\sim5$--6 orders of magnitude with respect to that at line centre, making it ideally suited to probing the order unity fluctuations in $\nf$ during the patchy EoR.

Detecting damping wing absorption in galaxy spectra generally requires large statistical samples, as well as assumptions about their redshift evolution and/or clustering properties \citep[e.g.][]{HaimanSpaans1999,Ouchi:2010p1,Stark:2010p1,Pentericci:2011p1,Ono:2012p1,Caruana:2014p1,Schenker:2014p1,Mesinger:2015p1584,SM15,Mason:2018}.
QSOs on the other hand are much rarer objects; however, they are also much brighter allowing a damping wing, if present, to be recovered from a single observed spectrum.

Using bright QSOs to constrain the EoR requires two key ingredients: (i) a knowledge of the intrinsic QSO emission; and (ii) a knowledge of the absorption caused by the EoR.
Both (i) and (ii) need to be estimated statistically, with the uncertainties carefully quantified, since we are relying on a single object to place constraints on the EoR.  We briefly discuss each in turn.

The intrinsic spectrum can be estimated from a composite of lower redshift objects \citep[e.g.][]{Francis:1991p5112,Brotherton:2001p1,VandenBerk:2001p3887,Telfer:2002p5713}.
However, a statistical reconstruction should take advantage of all of the available data from the QSO in question.  In particular, bright high-$z$ QSOs seem to exhibit anomalously large \civ{} blueshifts \citep{Mazzucchelli:2017}; as the \lya\ blue-shift is strongly correlated with the \civ\ blueshift, generic QSO templates are unlikely to fit the \lya\ lines of high-$z$ QSOs (e.g. \citealt{Bosman:2015p5005}).
In \citet{Greig:2017a}, we developed a reconstruction method which samples a covariance matrix of \lya\ and other strong emission line profiles from a sample of $\sim1700$ moderate-$z$ QSOs. This approach directly uses the strength and shape of the observed emission lines (e.g.\ \civ{}, \siv{} and \ciii{}) to recover the intrinsic \lya{} profile, with a statistical characterisation of the recovery.  Typical reconstruction errors are of order a few percent around the \lya\ line.

A common alternative approach is to deconstruct the QSO emission into principle component vectors, and fit these to the spectrum \citep[e.g.][]{Boroson:1992p4641,Francis:1992p5021,Suzuki:2005p5157,Suzuki:2006p4770,Lee:2011p1738,Paris:2011p4774}. \citet{Davies:2018a} recently introduced a sophisticated version of this principle component analysis (PCA), decomposing the QSO spectrum into components redward and blueward of $\lambda = 1280$\AA. They then reconstruct the \lya\ profile from the correlations between these red and blue side PCA components.  This mapping from (unattenuated) red-side components to the blue-side components, and the associated errors,  are trained on a large data set of $\sim$ 13,000 moderate-$z$ spectra.  They obtain $\sim$ percent level errors in the recovery similar to \citet{Greig:2017a}.

The second requirement for EoR constraints is a model for the attenuation by the EoR.  The EoR damping wing can attenuate the source flux both on the blue and the red side of the \lya\ line.  The source QSO is capable of highly ionising its environment, allowing some blue-side transmission to be seen (the so-called proximity zone).  Inside the proximity zone, the attenuation is a combination of: (i) resonant absorption by a fluctuating \lya\ forest, and (ii) a smooth damping wing from the more distant cosmic \hi\ patches.  Modelling (i) requires high-resolution simulations of the local QSO environment, while modelling (ii) requires ultra large-scale simulations of the EoR morphology (e.g. \citealt{Mesinger:2004p5737, Maselli:2007p5744, Bolton:2011p1063,Keating:2015p5004,Eilers:2017}).

In contrast, the attenuation on the red-side of the \lya\ line is free from resonant absorption, requiring only an understanding of large-scale EoR morphology to compute the associated damping wing absorption.  However, the damping wing imprint is weaker on the red side (far from the cosmic \hi\ patches), making it more degenerate with the intrinsic QSO emission.

In \citet{Greig:2017b} we combined the \lya\ reconstruction technique of \citet{Greig:2017a} with large-scale EoR simulations of \citet{Mesinger:2016p1}, constraining the hydrogen neutral fraction from the spectra of the $z=7.1$ QSO ULASJ1120+0641 \citep[hereafter J1120;][]{Mortlock:2011p1049}.  Our Bayesian framework  recovered strong evidence for an IGM damping wing redward of \lya\ ($\bar{x}_{\hi{}} = 0.40\substack{+0.21 \\ -0.19}$ at 68 per cent confidence). Subsequent analysis by \citet{Davies:2018b} found similar results, $\bar{x}_{\hi{}} = 0.48\substack{+0.26 \\ -0.26}$ at 68 per cent confidence.

In this work, we apply the same analysis framework to the spectrum of the recently-discovered $z=7.5$ QSO, ULASJ1342+0928 (hereafter J1342; \citealt{Banados:2018}).  Using their own analysis method, which performs the reconstruction using blue+red PCA components and models the proximity zone in addition to the red-side damping wing imprint, \citet{Davies:2018a}  finds $\bar{x}_{\hi{}} = 0.60\substack{+0.20 \\ -0.23}$ at 68 per cent confidence.  As the analysis methods of \citet{Greig:2017a} and \citet{Davies:2018a} are different (we go into more details below), this work, applied to the same input spectrum as \citet{Davies:2018a}, serves as an independent and complementary verification of the inferred EoR constraints from J1342.

This work is structured as follows. In Section~\ref{sec:Method} we briefly outline our analysis pipeline and in Section~\ref{sec:results} we provide our main results and discussion. In Section~\ref{sec:Conclusion} we finish with our closing remarks. Throughout we adopt the background cosmological parameters: ($\Omega_\Lambda$, $\Omega_{\rm M}$, $\Omega_b$, $n$, $\sigma_8$, $H_0$) = (0.69, 0.31, 0.048, 0.97, 0.81, 68 km s$^{-1}$ Mpc$^{-1}$), consistent with cosmic microwave background anisotropy measurements by the Planck satellite \citep{Collaboration:2015p4320} and unless otherwise stated, distances are quoted in comoving units.

\section{Method} \label{sec:Method}

\subsection{Reconstruction of the intrinsic \lya{} profile} \label{sec:Reconstruction}

In \citet{Greig:2017a}, we constructed a covariance matrix to characterise correlations between the emission line parameters\footnote{Each component of the emission line is modelled as a Gaussian, fully described by the peak height, width and velocity offset from systemic.} from the four most prominent high ionisation lines, \lya{}, \civ{}, \sioiv{} and \ciii{}. For both \lya{} and \civ{} we found a stronger preference for a broad and narrow component Gaussian to describe the line profile\footnote{Note that in the construction of this dataset we removed QSOs where the \lya{} line profile was not well characterised by two Gaussian components (see Appendix C in \citealt{Greig:2017a}). In most cases, the primary cause of this were absorption features at or near \lya{}. In total, this resulted in $\sim150$ QSOs being excluded from our final dataset (i.e. $\lesssim10$~per cent).}. Finally, we simultaneously fit a single power-law continuum. The dataset comprised 1673 moderate-$z$ ($2.08 < z < 2.5$)\footnote{This dataset uses the SDSS-III pipeline redshift to convert to rest-frame (see Appendix A of \citealt{Greig:2017a} for discussions on the redshift choice).}, high signal to noise (S/N $>15$) QSOs from SDSS-III (BOSS) DR12 \citep{Dawson:2013p5160,Alam:2015p5162}. 

With this covariance matrix, we then perform our reconstruction of the intrinsic \lya{} profile of J1342 as follows:
\begin{itemize}
\item We fit the rest-frame spectrum of J1342 at $\lambda > 1275$\AA\ using the [\cii{}] redshift \citep{Venemans:2017}\footnote{While we have used the [\cii{}] redshift for J1342, we do not have [\cii{}] redshifts for the SDSS-III dataset. This difference in redshift choice can lead to biases in the recovered line blueshifts from the line fitting and reconstruction pipelines. However, this bias is sub-dominant compared to the scatter in the correlations in the emission line parameters and variations between the different lines of sight through our EoR simulations.}, obtaining estimates of the continuum and the \sioiv{}, \civ{} and \ciii{} emission line profiles (we simultaneously fit for absorption lines by modelling each with a single Gaussian profile).
\item Using these red-side component fits, we collapse the 18-dimensional (Gaussian distributed) covariance matrix into a six dimensional estimate of the intrinsic \lya{} emission line profile (a two component Gaussian, each with an amplitude, width and velocity offset).
\item We then draw intrinsic \lya{} profiles from this distribution, applying a flux prior within the range $1250 < \lambda < 1275$\AA\ to ensure our reconstructed profiles fit the observed spectrum over this range.\footnote{Note, this prior range differs from the $1230 < \lambda < 1275$\AA\ used in \citet{Greig:2017b}. Here we are more conservative in our choice owing to the evidence of a stronger damping wing imprint that may extend beyond 1230\AA\ and the lower S/N of the observed spectrum. If we were to relax the prior range used for J1120 \citep{Greig:2017b}, the overall constraints would remain essentially unchanged, however the PDFs would be slightly broader.}
\end{itemize}

\subsection{The IGM damping wing during the EoR} \label{sec:IGMDampingWing}

\begin{figure*} 
	\begin{center}
	  \includegraphics[trim = 0.25cm 1.2cm 0cm 0.7cm, scale = 0.495]{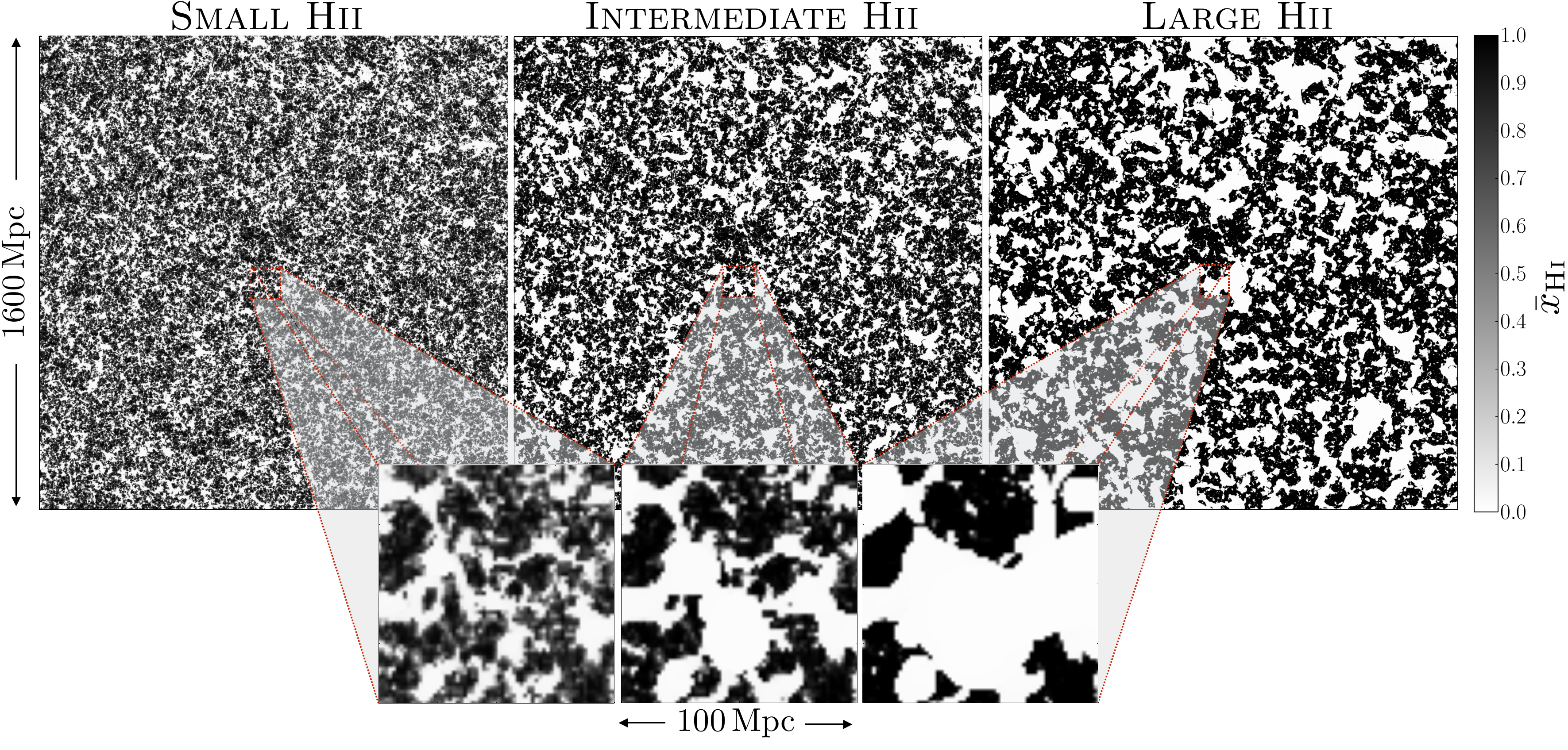}
	\end{center}
        \caption[]{
          Slices through the reionisation fields used in this study, taken at \avenf\ = 0.5.  The slices correspond to the \smallHII, \intermediateHII, and \largeHII\ models, from left to right (see text and \citealt{Mesinger:2016p1} for more details).  The large-scale (zoom-in) slice is 1600 (100) Mpc across, and 1.6 Mpc deep (one cell).
}
\label{fig:xHslices}
\end{figure*}

We compute our IGM damping wing profiles using the
Evolution of 21-cm Structure (EOS; \citealt{Mesinger:2016p1})\footnote{http://homepage.sns.it/mesinger/EOS.html} 2016 simulations. These comprise 1.6~Gpc on a side semi-numerical reionisation simulations on a 1024$^3$ grid, including state-of-the-art sub-grid prescriptions for inhomogeneous recombinations and photo-heating suppression of star-formation. We consider three different EoR morphologies, characterised by different efficiencies of star-formation inside low-mass halos, and visualised in Fig. \ref{fig:xHslices}:
\begin{itemize}
\item {\bf \smallHII} --  EoR driven by galaxies residing in $M_h \gsim 10^8 M_\odot$ haloes. In this scenario the EoR morphology is characterised by numerous small cosmic \hii{} regions.
\item {\bf \intermediateHII} --  EoR driven by galaxies residing in $M_h \gsim 10^9 M_\odot$ haloes. An intermediate scenario between the \smallHII\ and \largeHII\ models. We consider this to be our fiducial model\footnote{Although this is currently highly uncertain, here we motivate the \intermediateHII\ EoR morphology as a reasonable choice.  Some groups have recently suggested there might be weak evidence  of a turn-over starting to appear in the faint end of the lensed $z\sim6$ LFs (Yue et al. 2018; Atek et al. 2018).  Moreover, the current consensus of EoR observations prefers a late, moderately-extended reionisation, most naturally driven by galaxies of intermediate masses (e.g. Mitra et al. 2017; Fig. 11 in \citealt{Greig:2017EORhist}).}.
\item {\bf \largeHII} --  EoR driven by galaxies residing in $M_h \gsim 10^{10} M_\odot$ haloes, with an EoR morphology characterised by more spatially extended \hii{} structures.
\end{itemize}
We note that in \citet{Greig:2017b} we mistakenly represented the \intermediateHII\ model as the \largeHII\ model; however, the EoR morphology had limited impact on the results of that work, as we shall re-confirm below.

We extract a total of $10^5$ synthetic IGM damping wing profiles, constructed from 10 randomly oriented sightlines emanating from the centres of $10^4$ haloes between $6\times10^{11} < M_h < 3\times10^{12}$ $M_\odot$ at $z=7.5$. When computing the cumulative contribution from all encountered \hi{} patches, we exclude the first $\sim11$ comoving Mpc (1.3 physical Mpc) consistent with the estimated near-zone of J1342 \citep{Banados:2018}. The \igm{} neutral fraction at $z=7.5$ is then left as a free parameter by sampling the corresponding ionisation fields obtained from different redshift snapshots from the EoS simulations (i.e. different \avenf).

\subsection{Jointly fitting the IGM damping wing and intrinsic \lya{} profile distributions} \label{sec:JointFitting}

Finally, we infer the IGM neutral fraction by fitting the observed spectrum of J1342 by simultaneously sampling the distributions of both the intrinsic \lya{} line profile and the synthetic IGM damping wing profiles. Our procedure is as follows:
\begin{enumerate}
\item We draw $\sim10^5$ reconstructed \lya{} line profiles directly from the procedure outlined in Section~\ref{sec:Reconstruction}.
\item Each intrinsic profile is multiplied by the 10$^5$ synthetic damping wing opacities in Section~\ref{sec:IGMDampingWing}, to produce $\sim10^{10}$ mock spectra for each \avenf\ snapshot and EoR morphology.
\item All $\sim10^{10}$ mock spectra are then compared to the observed spectrum of J1342 over $1218$\AA $ < \lambda < 1230$\AA\ (consistent with \citealt{Greig:2017b}).
\item The resulting likelihood, averaged (i.e. marginalised) over all $\sim10^{10}$ mock spectra, is then assigned to that particular \avenf.
\item Steps (ii)--(iv) are then repeated for each \avenf\ to obtain a final 1D probability distribution function (PDF) of \avenf\ for each of the EoR morphologies.
\end{enumerate}

\section{Results and Discussion} \label{sec:results}

\begin{figure*} 
	\begin{center}
	  \includegraphics[trim = 0cm 1.2cm 0cm 0.8cm, scale = 0.425]{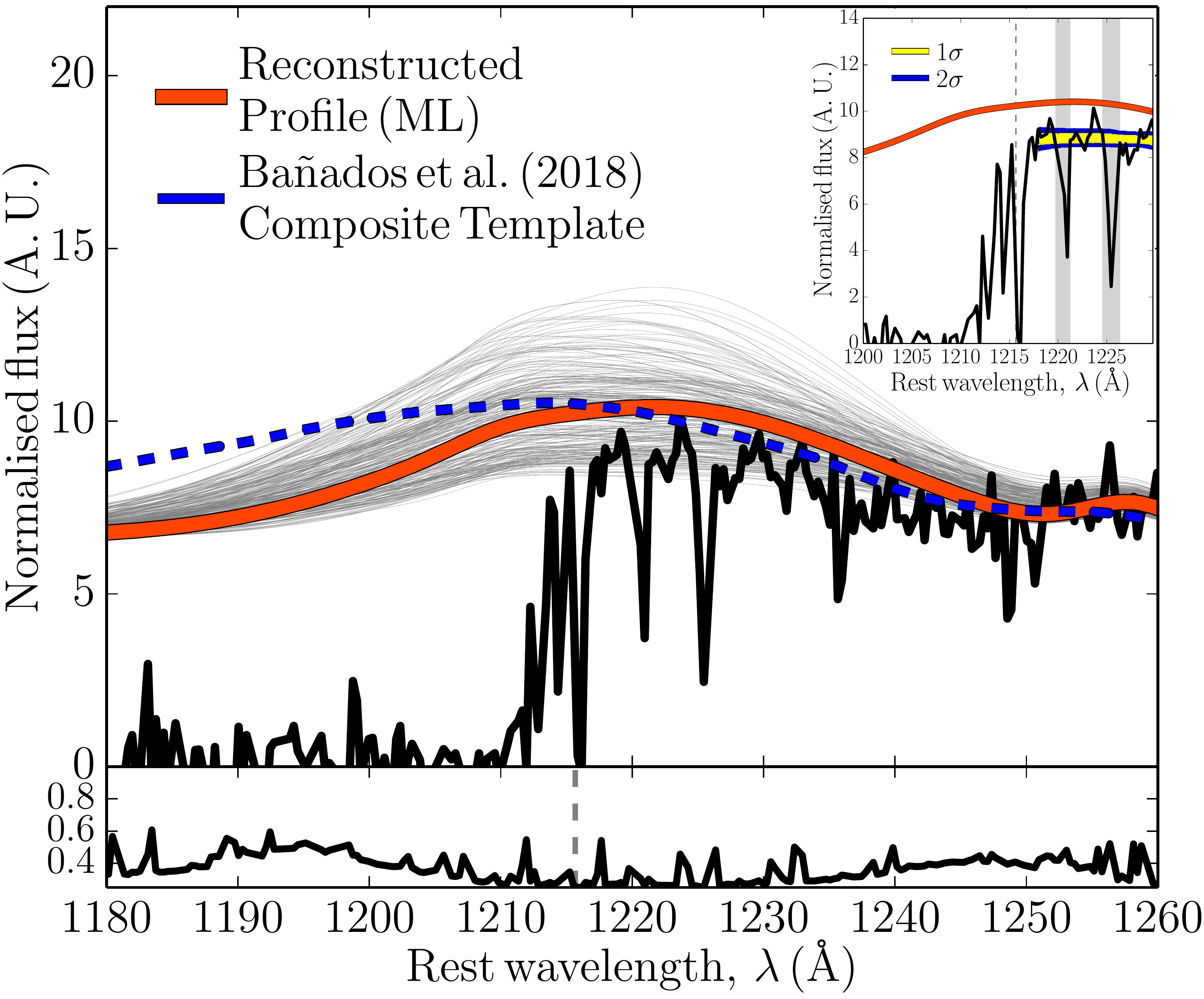}
	\end{center}
\caption[]{The reconstructed maximum likelihood \lya{} emission line profile (red curve; shown for visualisation purposes only) and a subsample of 300 \lya{} line profiles (thin grey curves) randomly drawn from the full posterior distribution of reconstructed profiles (see Section~\ref{sec:Reconstruction}; i.e.\ the analysis pipeline samples the full posterior rather than just the maximum likelihood profile). The black curve (and associated error below) corresponds to the combined Magellan/FIRE and Gemini/GNIRS spectrum, the blue dashed line corresponds to the QSO composite template constructed by \citet{Banados:2018} and the grey dashed line denotes rest-frame \lya{}. The zoom-in around \lya{} highlights the recovered imprint of the IGM damping wing profile with the yellow (blue) shaded region denoting the 68 (95) percentiles of total (intrinsic + damping wing absorption) flux over the $1218 < \lambda < 1230$\AA\ fitting region used in our analysis (see Section~\ref{sec:JointFitting}). Vertical grey shaded regions denote flux pixels masked from our analysis owing to strong absorption features.
}
\label{fig:Profile}
\end{figure*}

In Figure~\ref{fig:Profile} we provide our reconstructed intrinsic \lya{} emission line profile. The red curve corresponds to the best-fit (maximum likelihood; ML) reconstructed profile, while the 300 thin grey lines are posterior samples, illustrating the breadth of the uncertainties in the reconstruction pipeline.

Our reconstructed intrinsic \lya{} profile is almost entirely dominated by a single, broad component Gaussian. This arises owing to the strong preference for a large, broad component Gaussian to characterise the \civ{} emission line in J1342 and the corresponding correlation between the \lya--\civ\ broad components.
An extremely small, narrow component Gaussian can be identified near $1210$\AA\ owing to the extreme \civ{} blueshift of J1342 ($\sim6600$~km/s). Though $z>6.5$ QSOs typically exhibit extreme \civ{} blueshifts \citep{Mazzucchelli:2017}, J1342 itself is an outlier with a blueshift more than a factor of two larger than J1120. The covariance matrix developed by \citet{Greig:2017a} does not include QSOs with such  extreme \civ{} blueshifts as J1342. However, in \citet{Greig:2017b} (see Fig. A1 and the associated discussion) we verified that the covariance matrix of emission line properties could be extrapolated to reconstruct QSOs within our dataset with \civ{} blueshifts similar to J1120 ($\sim2000$~km/s). Since the extrapolation works for J1120, we assume that the extrapolation is equally valid for J1342.

Our recovered intrinsic \lya{} profile for J1342 is similar to the SDSS/BOSS composite template constructed by \citet{Banados:2018} (blue-dashed curve). Their composite was constructed from 46 QSOs with similar \civ{} blueshifts relative to \mgii{} and \civ{} equivalent widths of J1342. 
Likewise, the \citet{Davies:2018a} PCA-based reconstruction exhibits a qualitatively similar ML profile. Specifically, their reconstruction prefers a dominant contribution from a broad component-like feature for the \lya\ profile, with a secondary narrower component near $1210$\AA. However, the distribution of reconstructed profiles differs in our two approaches.  Their PCA method does not provide a direct estimate of the associated uncertainty in the fit.  To estimate the uncertainty and be able to forward model the intrinsic emission, they construct a covariance matrix of fit errors for each spectral bin, by performing a reconstruction on their SDSS sample and comparing to the actual spectra.
As a result, their profile samples (c.f. the thin blue curves of their figure 8) have unphysical oscillatory features (though it is likely such features average out in their full analysis).
Moreover, our reconstruction method prefers a notably broader distribution of reconstructed \lya\ profiles than that presented by \citet{Davies:2018b}. This broad scatter arises from the correlations amongst individual emission line profiles, which can have notable scatter \citep{Greig:2017a}\footnote{It is also broader than the distribution shown for J1120 \citep{Greig:2017b} where we presented only the 68 percentiles while also using a more aggressive flux prior.}. The large spread in reconstructed intrinsic emission profiles translates to a broader PDF for the inferred IGM neutral fraction (discussed below).

In the zoom-in panel of Figure~\ref{fig:Profile}, we present the confidence intervals from our joint fitting of the IGM damping wing and reconstructed \lya\ profiles (Section~\ref{sec:JointFitting}). Here, the yellow (blue) shaded regions correspond to the 68 (95) percentiles, using the \intermediateHII\ EoR morphology.
The red curve is the same best-fit reconstructed profile as shown in the left panel, and is used to illustrate the impact of the IGM damping wing on the intrinsic \lya{} profile.  The offset of the red curve and the yellow/blue strips is suggestive of the presence of an IGM damping wing; however, the spread around the ML shown with the grey curves in main panel makes such evidence weak.

\begin{figure}
	\begin{center}
		\includegraphics[trim = 0cm 0.8cm 0cm 0.2cm, scale = 0.42]{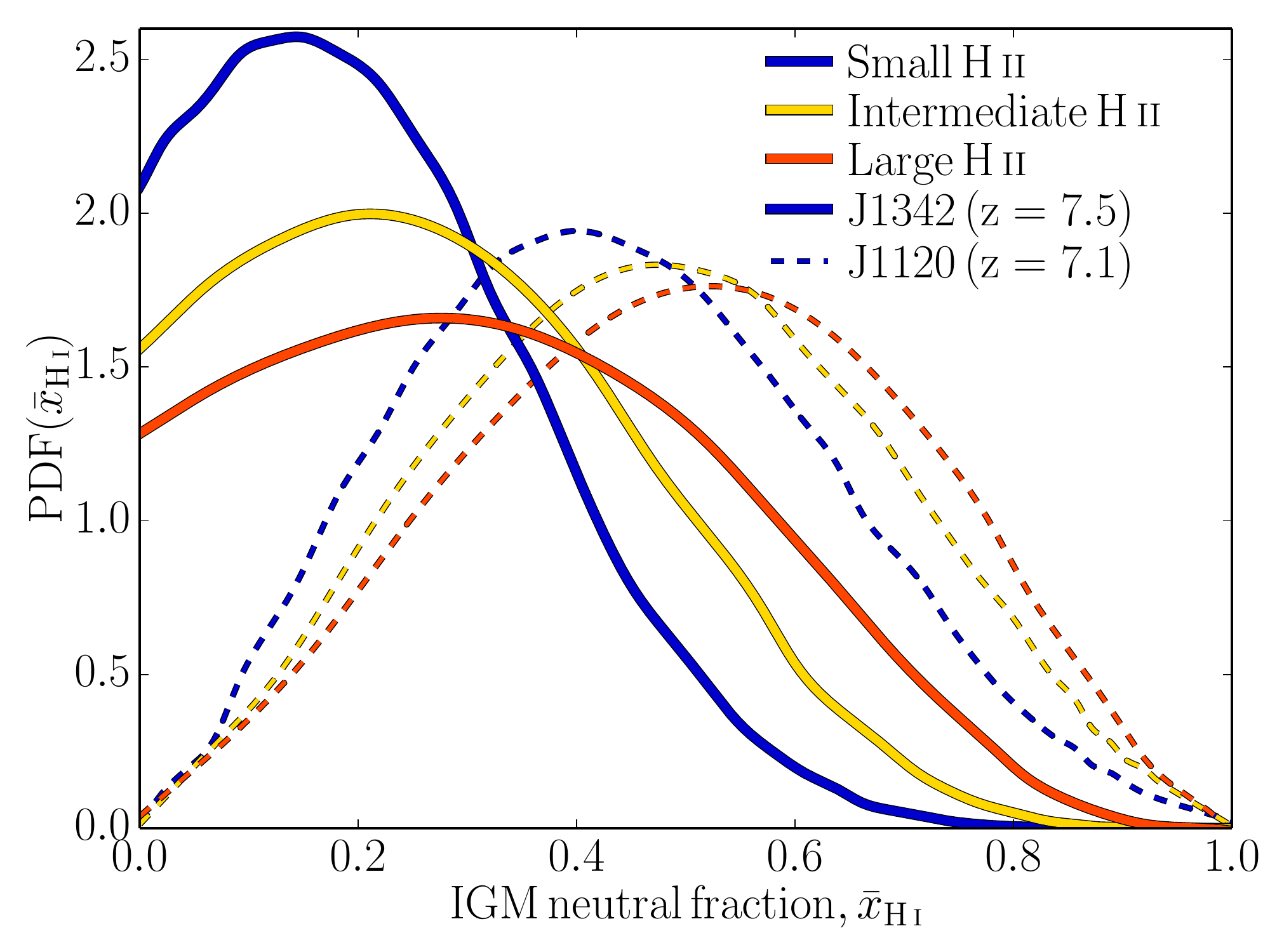}
	\end{center}
\caption[]{The marginalised 1D PDF of the IGM neutral fraction for the three different EoR morphologies (see Section~\ref{sec:IGMDampingWing}): \smallHII\ (blue curve), \intermediateHII\ (yellow curve) and \largeHII\ (red curve).
}
\label{fig:PDFs}
\end{figure}

We quantify this in Figure~\ref{fig:PDFs}, which presents the main results of this work: the 1D PDFs of the IGM neutral fraction for the three different EoR morphologies.
In summary, for each EoR morphology we find:
\begin{itemize}
\item {\bf \smallHII}; $\bar{x}_{\hi{}}\sim0.14$, $\bar{x}_{\hi{}} < 0.28$ ($0.51$) at 68 (95) per cent
\item {\bf \intermediateHII}; $\bar{x}_{\hi{}} = 0.21\substack{+0.17 \\ -0.19}$ (68 per cent), $\bar{x}_{\hi{}} < 0.61$ (95 per cent)
\item {\bf \largeHII}; $\bar{x}_{\hi{}} = 0.28\substack{+0.20 \\ -0.23}$ (68 per cent), $\bar{x}_{\hi{}} < 0.70$ (95 per cent).
\end{itemize}

We do not find strong evidence for J1342 to be in a significantly neutral IGM. Depending on the EoR model, the spectrum is consistent with being in a fully-ionised Universe at $\sim$ 1--2$\sigma$.  This broad distribution is driven in part by the afore-mentioned large scatter in the reconstruction.

Similar to, but slightly stronger than J1120 (also shown in the plot; \citealt{Greig:2017b}), we find a weak EoR morphological dependence on the recovered IGM neutral fraction. Compared to J1120, the PDFs are closer to low \avenf\ values.  The sightline-to-sightline scatter in damping wing opacity increases with decreasing \avenf\ (see e.g. figure 3 in \citealt{Mesinger:2008p5748}), driving broader PDFs.  In this regime, the absorption is more sensitive to the incidence with the remaining rare neutral patches whose sizes and separation depend on the source model.

Although at first glance it might seem strange that the neutral fraction at $z=7.5$ preferred by J1342 is lower than the one preferred by J1120 at $z=7.1$, it is important to note that the distributions are quite broad.  Thus a physically reasonable neutral fraction which evolves monotonically with redshift is perfectly consistent within the errors.  Specifically, comparing to the EoR history constraints in Fig. 10 of \citet{Greig:2017EORhist}, we see that the \intermediateHII\ model constraint of $\bar{x}_{\hi{}} = 0.21\substack{+0.17 \\ -0.19}$ (68 per cent) falls comfortably within the 1 $\sigma$ range at $z=7.5$.

Our results are in mild tension ($\sim$ 1 -- 1.5$\sigma$) with the constraints in \citet{Banados:2018} and \citet{Davies:2018b}. These authors find stronger evidence of an incomplete reionisation, $\bar{x}_{\hi{}} = 0.56\substack{+0.21 \\ -0.18}$ \citep[Model A;][]{Banados:2018} and $\bar{x}_{\hi{}} = 0.60\substack{+0.20 \\ -0.23}$ \citep{Davies:2018a}. While it is difficult to do a direct comparison with their works, we can speculate on the main causes for this difference.  The two approaches yield different results for both components in the analysis: (i) the reconstruction of the intrinsic emission profile; and (ii) the attenuation from the IGM.  We discuss these briefly in turn.

Although qualitatively similar, the PCA reconstruction in \citet{Davies:2018b} results in a somewhat higher amplitude intrinsic emission than we predict with our emission line covariance approach (see Figure~\ref{fig:Profile}). This would naturally require stronger attenuation of the intrinsic flux to achieve a fit to the observed spectrum. As a result, higher neutral fractions are preferred.  The opposite is true for the reconstructions of J1120 by these authors.
Furthermore, as mentioned previously the scatter around the ML is smaller in their reconstruction, resulting in a narrower PDFs of the inferred neutral fraction.

The main difference in modelling the IGM attenuation lies in the treatment of the near zone transmission.  Our analysis ignores the flux in the near zone, fitting only the damping wing redward of the line centre (and any probable infall). In contrast, the state-of-the-art approach of \citet{Davies:2018b} also uses the near zone flux of J1120 and J1342 when comparing to simulated spectra.  They do this by performing a 1D radiative transfer through a 100 $h^{-1}$ Mpc \lya\ forest simulation, adding to this a smooth damping contribution from a large semi-numerical simulation of the EoR. Their radiative transfer assumes a constant ionising QSO luminosity, which is on for a fixed time; their final results marginalise over this quasar lifetime.\footnote{As pointed out in \citet{Davies:2018b}, our analysis effectively assumes a complicated prior over such a quasar lifetime.  By ensuring that the surrounding \hii\ regions in our EoR simulations are {\it at least} as large as the observed near zone, we are essentially assuming a minimum QSO contribution.  For the damping wing redward of \lya\ used in our analysis, this is mainly relevant in very neutral universes in which galaxies could not by themselves carve out large enough \hii\ bubbles surrounding the QSO.  As a result, our analysis is slightly biased against very neutral Universes, what we call a ``conservative'' choice in \citet{Greig:2017b}.  In any case, \citet{Davies:2018b} show that the assumed QSO lifetime has a very negligible impact on their results for a reasonable range of values $10^4$ -- $10^7$ yr (e.g. see their figures 7 and 10).}
Using the resulting near zone models when comparing to the observed spectra adds constraining power, but makes the results much more model dependent.  For example, the \lya\ forest simulations used in that work have a volume which is roughly a factor of $\gsim$~300 too small to capture the rare, biased $\sim10^{12} \Msun$ halos expected to host these QSOs. As a result, they do not simulate the biased environment of these QSOs, which might have important consequences for the corresponding near zone transmission profiles.

Finally we comment on the quality of the observed spectrum. In all works, the QSO spectrum used in the analysis has been the combined Magellan/FIRE and Gemini/GNIRS spectrum, which corresponds to a resolution of $R\sim1800$ \citep{Banados:2018}. This relatively low S/N spectrum results in numerous spurious features in the emission spectrum which can hinder attempts to characterise the QSO continuum or to accurately fit the various emission lines, a prerequisite for this work and the PCA approach of \citet{Davies:2018a}. Further, features in emission (absorption) could artificially bias recovered IGM neutral fractions to lower (higher) values. Although attempts have been made to identify and mask problematic regions of the spectrum, this is made difficult owing to the lower resolution. It will therefore be fruitful to return to this analysis once a deeper spectrum is obtained.

\section{Conclusion} \label{sec:Conclusion}

With the recent detection of the $z=7.5$ QSO, J1342 \citep{Banados:2018}, we perform an independent analysis quantifying the damping wing imprint from the EoR. \citet{Banados:2018} and \citet{Davies:2018b} have already analysed this source using their own analysis pipelines recovering $\bar{x}_{\hi{}} \sim 0.6$. In both previous works, the red and blue side of the \lya\ line is used for the constraints (e.g. Models B and C of \citealt{Banados:2018}).  Here we focus only on the
red side of the line ($>1218$\AA).  This is a conservative choice in that it is less constraining but more model independent as it is not sensitive to the complicated modelling of the near zone transmission.

We use the same analysis pipeline that was applied to the $z=7.1$ QSO, J1120 \citep{Greig:2017b}. We perform a reconstruction of the intrinsic (unattenuated) QSO profile near \lya{} using a covariance matrix of correlations between various known emission lines \citep{Greig:2017a} and then couple these with synthetic IGM damping wing profiles extracted from large EoR simulations with different morphologies. We then fit $\sim10^{10}$ template profiles to the observed spectrum of J1342 between 1218 -- 1230\AA\ within a Bayesian framework. We recover systematically lower values that those presented by \citet{Banados:2018} and \citet{Davies:2018b}, although they are consistent at $\sim$ 1 -- 1.5 $\sigma$.  Specifically, we find for our three EoR morphologies:
\begin{itemize}
\item {\bf \smallHII}; $\bar{x}_{\hi{}}\sim0.14$, $\bar{x}_{\hi{}} < 0.28$ ($0.51$) at 68 (95) per cent
\item {\bf \intermediateHII}; $\bar{x}_{\hi{}} = 0.21\substack{+0.17 \\ -0.19}$ (68 per cent), $\bar{x}_{\hi{}} < 0.61$ (95 per cent)
\item {\bf \largeHII}; $\bar{x}_{\hi{}} = 0.28\substack{+0.20 \\ -0.23}$ (68 per cent), $\bar{x}_{\hi{}} < 0.70$ (95 per cent).
\end{itemize}
We suspect the primary differences arise from the reconstruction of the intrinsic QSO profile and the modelling of the host QSO environment. Our results are consistent within 1 $\sigma$ with previous estimates of the global EoR history, which are suggestive of a moderately-extended reionisation.  

\section*{Acknowledgements}

We thank Fred Davies for comments on a draft version of this manuscript.
Parts of this research were supported by the Australian Research Council Centre of Excellence for All Sky Astrophysics in 3 Dimensions (ASTRO 3D), through project number CE170100013. AM acknowledges funding support from the European Research Council (ERC) under the European Union's Horizon 2020 research and innovation programme (grant agreement No 638809 -- AIDA -- PI: AM).

\bibliography{Papers}

\begin{thebibliography}{44}
\expandafter\ifx\csname natexlab\endcsname\relax\def\natexlab#1{#1}\fi

\bibitem[{Alam {et~al.}(2015)}]{Alam:2015p5162}
Alam S., {et~al.}, 2015, ApJS, 219, 12

\bibitem[{Ba{\~n}ados {et~al.}(2018)}]{Banados:2018}
Ba{\~n}ados E., {et~al.}, 2018, Nature, 553, 473

\bibitem[{Bolton {et~al.}(2011)Bolton, Haehnelt, Warren, Hewett, Mortlock,
  Venemans, McMahon, \& Simpson}]{Bolton:2011p1063}
Bolton J.~S., Haehnelt M.~G., Warren S.~J., Hewett P.~C., Mortlock D.~J.,
  Venemans B.~P., McMahon R.~G., Simpson C., 2011, MNRAS, 416, L70

\bibitem[{Boroson \& Green(1992)}]{Boroson:1992p4641}
Boroson T.~A., Green R.~F., 1992, ApJS, 80, 109

\bibitem[{Bosman \& Becker(2015)}]{Bosman:2015p5005}
Bosman S. E.~I., Becker G.~D., 2015, MNRAS, 452, 1105

\bibitem[{Brotherton {et~al.}(2001)Brotherton, Tran, Becker, Gregg,
  Laurent-Muehleisen, \& White}]{Brotherton:2001p1}
Brotherton M.~S., Tran H.~D., Becker R.~H., Gregg M.~D., Laurent-Muehleisen
  S.~A., White R.~L., 2001, ApJ, 546, 775

\bibitem[{Caruana {et~al.}(2014)Caruana, Bunker, Wilkins, Stanway, Lorenzoni,
  Jarvis, \& Ebert}]{Caruana:2014p1}
Caruana J., Bunker A.~J., Wilkins S.~M., Stanway E.~R., Lorenzoni S., Jarvis
  M.~J., Ebert H., 2014, MNRAS, 442, 2831

\bibitem[{Davies {et~al.}(2018{\natexlab{a}})}]{Davies:2018a}
Davies F.~B., {et~al.}, 2018{\natexlab{a}}, preprint (arXiv:1801.07679)

\bibitem[{Davies {et~al.}(2018{\natexlab{b}})}]{Davies:2018b}
---, 2018{\natexlab{b}}, preprint (arXiv:1802.06066)

\bibitem[{Dawson {et~al.}(2013)}]{Dawson:2013p5160}
Dawson K.~S., {et~al.}, 2013, AJ, 145, 10

\bibitem[{Eilers {et~al.}(2017)Eilers, Davies, Hennawi, Prochaska, Luki{\'c},
  \& Mazzucchelli}]{Eilers:2017}
Eilers A.-C., Davies F.~B., Hennawi J.~F., Prochaska J.~X., Luki{\'c} Z.,
  Mazzucchelli C., 2017, ApJ, 840, 24

\bibitem[{Fan {et~al.}(2006)}]{Fan:2006p4005}
Fan X., {et~al.}, 2006, AJ, 132, 117

\bibitem[{Francis {et~al.}(1992)Francis, Hewett, Foltz, \&
  Chaffee}]{Francis:1992p5021}
Francis P.~J., Hewett P.~C., Foltz C.~B., Chaffee F.~H., 1992, ApJ, 398, 476

\bibitem[{Francis {et~al.}(1991)Francis, Hewett, Foltz, Chaffee, Weymann, \&
  Morris}]{Francis:1991p5112}
Francis P.~J., Hewett P.~C., Foltz C.~B., Chaffee F.~H., Weymann R.~J., Morris
  S.~L., 1991, ApJ, 373, 465

\bibitem[{George {et~al.}(2015)}]{George:2015p5869}
George E.~M., {et~al.}, 2015, ApJ, 799, 177

\bibitem[{Greig \& Mesinger(2017)}]{Greig:2017EORhist}
Greig B., Mesinger A., 2017, MNRAS, 465, 4838

\bibitem[{Greig {et~al.}(2017{\natexlab{a}})Greig, Mesinger, Haiman, \&
  Simcoe}]{Greig:2017b}
Greig B., Mesinger A., Haiman Z., Simcoe R.~A., 2017{\natexlab{a}}, MNRAS, 466,
  4239

\bibitem[{Greig {et~al.}(2017{\natexlab{b}})Greig, Mesinger, McGreer,
  Gallerani, \& Haiman}]{Greig:2017a}
Greig B., Mesinger A., McGreer I.~D., Gallerani S., Haiman Z.,
  2017{\natexlab{b}}, MNRAS, 466, 1814

\bibitem[{{Haiman} \& {Spaans}(1999)}]{HaimanSpaans1999}
{Haiman} Z., {Spaans} M., 1999, ApJ, 518, 138

\bibitem[{Keating {et~al.}(2015)Keating, Haehnelt, Cantalupo, \&
  Puchwein}]{Keating:2015p5004}
Keating L.~C., Haehnelt M.~G., Cantalupo S., Puchwein E., 2015, MNRAS, 454, 681

\bibitem[{Lee \& Spergel(2011)}]{Lee:2011p1738}
Lee K.-G., Spergel D.~N., 2011, ApJ, 734, 21

\bibitem[{Maselli {et~al.}(2007)Maselli, Gallerani, Ferrara, \&
  Choudhury}]{Maselli:2007p5744}
Maselli A., Gallerani S., Ferrara A., Choudhury T.~R., 2007, MNRAS, 376, L34

\bibitem[{Mason {et~al.}(2018)Mason, Treu, Dijkstra, Mesinger, Trenti,
  Pentericci, de~Barros, \& Vanzella}]{Mason:2018}
Mason C.~A., Treu T., Dijkstra M., Mesinger A., Trenti M., Pentericci L.,
  de~Barros S., Vanzella E., 2018, ApJ, 856, 2

\bibitem[{Mazzucchelli {et~al.}(2017)}]{Mazzucchelli:2017}
Mazzucchelli C., {et~al.}, 2017, ApJ, 849, 91

\bibitem[{Mesinger {et~al.}(2015)Mesinger, Aykutalp, Vanzella, Pentericci,
  Ferrara, \& Dijkstra}]{Mesinger:2015p1584}
Mesinger A., Aykutalp A., Vanzella E., Pentericci L., Ferrara A., Dijkstra M.,
  2015, MNRAS, 446, 566

\bibitem[{Mesinger \& Furlanetto(2008)}]{Mesinger:2008p5748}
Mesinger A., Furlanetto S.~R., 2008, MNRAS, 386, 1990

\bibitem[{Mesinger {et~al.}(2016)Mesinger, Greig, \&
  Sobacchi}]{Mesinger:2016p1}
Mesinger A., Greig B., Sobacchi E., 2016, MNRAS, 459, 2342

\bibitem[{Mesinger {et~al.}(2004)Mesinger, Haiman, \& Cen}]{Mesinger:2004p5737}
Mesinger A., Haiman Z., Cen R., 2004, ApJ, 613, 23

\bibitem[{Miralda-Escud{\'e}(1998)}]{MiraldaEscude:1998p1041}
Miralda-Escud{\'e} J., 1998, ApJ, 501, 15

\bibitem[{Mortlock {et~al.}(2011)}]{Mortlock:2011p1049}
Mortlock D.~J., {et~al.}, 2011, Nature, 474, 616

\bibitem[{Ono {et~al.}(2012)}]{Ono:2012p1}
Ono Y., {et~al.}, 2012, ApJ, 744, 83

\bibitem[{Ouchi {et~al.}(2010)}]{Ouchi:2010p1}
Ouchi M., {et~al.}, 2010, ApJ, 723, 869

\bibitem[{P{\^a}ris {et~al.}(2011)}]{Paris:2011p4774}
P{\^a}ris I., {et~al.}, 2011, A\&A, 530, A50

\bibitem[{Pentericci(2011)}]{Pentericci:2011p1}
Pentericci L., 2011, ApJ, 443, 132

\bibitem[{{Planck Collaboration XIII}(2016)}]{Collaboration:2015p4320}
{Planck Collaboration XIII}, 2016, A\&A, 594, 13

\bibitem[{{Rybicki} \& {Lightman}(1979)}]{Rybicki1979}
{Rybicki} G.~B., {Lightman} A.~P., 1979, {Radiative Processes in Astrophysics,
  Wiley-Interscience, New York.}

\bibitem[{Schenker {et~al.}(2014)Schenker, Ellis, Konidaris, \&
  Stark}]{Schenker:2014p1}
Schenker M.~A., Ellis R.~S., Konidaris N.~P., Stark D.~P., 2014, ApJ, 795, 20

\bibitem[{{Sobacchi} \& {Mesinger}(2015)}]{SM15}
{Sobacchi} E., {Mesinger} A., 2015, MNRAS, 453, 1843

\bibitem[{Stark {et~al.}(2010)Stark, Ellis, Chiu, Ouchi, \&
  Bunker}]{Stark:2010p1}
Stark D.~P., Ellis R.~S., Chiu K., Ouchi M., Bunker A., 2010, MNRAS, 408, 1628

\bibitem[{Suzuki(2006)}]{Suzuki:2006p4770}
Suzuki N., 2006, ApJS, 163, 110

\bibitem[{Suzuki {et~al.}(2005)Suzuki, Tytler, Kirkman, O'Meara, \&
  Lubin}]{Suzuki:2005p5157}
Suzuki N., Tytler D., Kirkman D., O'Meara J.~M., Lubin D., 2005, ApJ, 618, 592

\bibitem[{Telfer {et~al.}(2002)Telfer, Zheng, Kriss, \&
  Davidsen}]{Telfer:2002p5713}
Telfer R.~C., Zheng W., Kriss G.~A., Davidsen A.~F., 2002, ApJ, 565, 773

\bibitem[{{Vanden Berk} {et~al.}(2001)}]{VandenBerk:2001p3887}
{Vanden Berk} D.~E., {et~al.}, 2001, AJ, 122, 549

\bibitem[{Venemans {et~al.}(2017)}]{Venemans:2017}
Venemans B.~P., {et~al.}, 2017, ApJL, 851, L8

\end{thebibliography}

\end{document}